%% file: 1331.tex
\documentstyle[emulateapj,rotate,times]{article}
 
\newcommand{\etal}{et~al.\ }
\newcommand{\PVdblt}{{\rm P}\kern 0.1em{\sc v}~$\lambda\lambda 1117, 1128$}
\newcommand{\CaIIdblt}{{\rm Ca}\kern 0.1em{\sc ii}~$\lambda\lambda 3934, 3969$}
\newcommand{\AlIIIdblt}{{\rm Al}\kern 0.1em{\sc iii}~$\lambda\lambda 1854, 1862$}
\newcommand{\CIVdblt}{{\rm C}\kern 0.1em{\sc iv}~$\lambda\lambda 1548, 1550$}
\newcommand{\MgIIdblt}{{\rm Mg}\kern 0.1em{\sc ii}~$\lambda\lambda 2796, 2803$}
\newcommand{\NVdblt}{{\rm N}\kern 0.1em{\sc v}~$\lambda\lambda 1238, 1242$}  
\newcommand{\SVIdblt}{{\rm S}\kern 0.1em{\sc vi}~$\lambda\lambda 933, 944$} 
\newcommand{\OVIdblt}{{\rm O}\kern 0.1em{\sc vi}~$\lambda\lambda 1031, 1037$} 
\newcommand{\SiIIdblt}{{\rm Si}\kern 0.1em{\sc ii}~$\lambda\lambda 1190, 1193$} 
\newcommand{\SiIVdblt}{{\rm Si}\kern 0.1em{\sc iv}~$\lambda\lambda 1393, 1402$} 
\newcommand{\PV}{\hbox{{\rm P}\kern 0.1em{\sc v}}}
\newcommand{\AlI}{\hbox{{\rm Al}\kern 0.1em{\sc i}}}
\newcommand{\AlII}{\hbox{{\rm Al}\kern 0.1em{\sc ii}}}
\newcommand{\AlIII}{{\hbox{\rm Al}\kern 0.1em{\sc iii}}}
\newcommand{\CaII}{\hbox{{\rm Ca}\kern 0.1em{\sc ii}}}
\newcommand{\CII}{\hbox{{\rm C}\kern 0.1em{\sc ii}}}
\newcommand{\CIIe}{\hbox{{\rm C$^{\ast}$}\kern 0.1em{\sc ii}}}
\newcommand{\CIII}{\hbox{{\rm C}\kern 0.1em{\sc iii}}}
\newcommand{\CIV}{\hbox{{\rm C}\kern 0.1em{\sc iv}}}
\newcommand{\CV}{\hbox{{\rm C}\kern 0.1em{\sc v}}}
\newcommand{\HI}{\hbox{{\rm H}\kern 0.1em{\sc i}}}
\newcommand{\HII}{\hbox{{\rm H}\kern 0.1em{\sc ii}}}
\newcommand{\Lya}{\hbox{{\rm Ly}\kern 0.1em$\alpha$}}
\newcommand{\Lyb}{\hbox{{\rm Ly}\kern 0.1em$\beta$}}
\newcommand{\Lyg}{\hbox{{\rm Ly}\kern 0.1em$\gamma$}}
\newcommand{\Lyd}{\hbox{{\rm Ly}\kern 0.1em$\delta$}}
\newcommand{\Lye}{\hbox{{\rm Ly}\kern 0.1em$\epsilon$}}
\newcommand{\Lyphi}{\hbox{{\rm Ly}\kern 0.1em$\phi$}}
\newcommand{\Lyfive}{\hbox{{\rm Ly}\kern 0.1em$5$}}
\newcommand{\Lysix}{\hbox{{\rm Ly}\kern 0.1em$6$}}
\newcommand{\Lyseven}{\hbox{{\rm Ly}\kern 0.1em$7$}}
\newcommand{\Lyeight}{\hbox{{\rm Ly}\kern 0.1em$8$}}
\newcommand{\Lynine}{\hbox{{\rm Ly}\kern 0.1em$9$}}
\newcommand{\Lyten}{\hbox{{\rm Ly}\kern 0.1em$10$}}
\newcommand{\Lyeleven}{\hbox{{\rm Ly}\kern 0.1em$11$}}
\newcommand{\HeI}{\hbox{{\rm He}\kern 0.1em{\sc i}}}
\newcommand{\HeII}{\hbox{{\rm He}\kern 0.1em{\sc ii}}}
\newcommand{\FeI}{\hbox{{\rm Fe}\kern 0.1em{\sc i}}}
\newcommand{\FeII}{\hbox{{\rm Fe}\kern 0.1em{\sc ii}}}
\newcommand{\FeIII}{\hbox{{\rm Fe}\kern 0.1em{\sc iii}}}
\newcommand{\MnII}{\hbox{{\rm Mn}\kern 0.1em{\sc ii}}}
\newcommand{\MgI}{\hbox{{\rm Mg}\kern 0.1em{\sc i}}}
\newcommand{\MgII}{\hbox{{\rm Mg}\kern 0.1em{\sc ii}}}
\newcommand{\MgIII}{\hbox{{\rm Mg}\kern 0.1em{\sc iii}}}
\newcommand{\NI}{\hbox{{\rm N}\kern 0.1em{\sc i}}}
\newcommand{\NII}{\hbox{{\rm N}\kern 0.1em{\sc ii}}}
\newcommand{\NIII}{\hbox{{\rm N}\kern 0.1em{\sc iii}}}
\newcommand{\NV}{\hbox{{\rm N}\kern 0.1em{\sc v}}}
\newcommand{\OVI}{\hbox{{\rm O}\kern 0.1em{\sc vi}}}
\newcommand{\OI}{\hbox{{\rm O}\kern 0.1em{\sc i}}}
\newcommand{\OII}{\hbox{[{\rm O}\kern 0.1em{\sc ii}]}}
\newcommand{\OIV}{\hbox{{\rm O}\kern 0.1em{\sc iv}]}}
\newcommand{\SI}{{\rm S}\kern 0.1em{\sc i}}
\newcommand{\SIV}{{\rm S}\kern 0.1em{\sc iv}}
\newcommand{\SVI}{{\rm S}\kern 0.1em{\sc vi}}
\newcommand{\SiI}{\hbox{{\rm Si}\kern 0.1em{\sc i}}}
\newcommand{\SiII}{\hbox{{\rm Si}\kern 0.1em{\sc ii}}}
\newcommand{\SiIII}{\hbox{{\rm Si}\kern 0.1em{\sc iii}}}
\newcommand{\SiIV}{\hbox{{\rm Si}\kern 0.1em{\sc iv}}}
\newcommand{\SII}{\hbox{{\rm S}\kern 0.1em{\sc ii}}}
\newcommand{\SIII}{\hbox{{\rm S}\kern 0.1em{\sc iii}}}
\newcommand{\NaI}{\hbox{{\rm Na}\kern 0.1em{\sc i}}}
\newcommand{\TiII}{\hbox{{\rm Ti}\kern 0.1em{\sc ii}}}
\newcommand{\kms}{\hbox{km~s$^{-1}$}}
\newcommand{\cmsq}{\hbox{cm$^{-2}$}}
\newcommand{\cc}{\hbox{cm$^{-3}$}}
\newcommand{\h}{\hbox{h$^{-1}$}}
\newcommand{\kmsM}{\hbox{km~s$^{-1}$~Mpc$^{-1}$}}
 
\tighten
\begin{document}
 
 
\lefthead{BOND ET~AL.}
\righthead{EXPANDING SUPERBUBBLES IN Q1331+170}


\title{Evidence for Expanding Superbubbles in a Galaxy at \lowercase{z}=0.7443\altaffilmark{1}}

\author{Nicholas~A.~Bond, Christopher~W.~Churchill, Jane~C.~Charlton\altaffilmark{2}} 
\affil{Department of Astronomy and Astrophysics \\
       The Pennsylvania State University \\ University Park, PA 16802
       \\ {\it bond, cwc, charlton@astro.psu.edu}}
\and 
\author{Steven S. Vogt}
\affil{UCO/Lick Observatories \\
       Board of Studies in Astronomy and Astrophysics \\
       University of California, Santa Cruz, CA 96054 \\
       {\it vogt@ucolick.org}}

\altaffiltext{1}{Based in part on observations obtained at the
W.~M. Keck Observatory, which is operated as a scientific partnership
among Caltech, the University of California, and NASA. The Observatory
was made possible by the generous financial support of the W.~M. Keck
Foundation.}
\altaffiltext{2}{Center for Gravitational Physics and Geometry}

\begin{abstract}
The intervening $z=0.7443$ {\MgII} absorption system in the spectrum
of MC~$1331+170$ shows an unusual series of line pairs, each with $\Delta v
\sim 30$~{\kms}.  These lines could be explained as the shells of
expanding superbubbles residing in the outer regions of an edge--on
spiral galaxy visible in the optical image of the MC~$1331+170$ field.
The color and brightness of this galaxy make it the most likely
candidate $z=0.7443$ absorber, though two other galaxies in the quasar
field could also be contributing to the {\MgII} absorption profile.
Kinematic models of absorption from compact groups and galaxy pairs
produce profiles largely inconsistent with the observed {\MgII}
spectrum.  Superbubbles would naturally generate more regular
structures such as those observed.  Photoionization models of the
superbubble shell are consistent with the observed profile for many
realistic physical conditions.  In a pure superbubble model, the large
velocity spread of the {\MgII} absorption system is inconsistent with
the expected spread of a quiescent, rotating disk.  This requires
unusual kinematics within the host galaxy, perhaps due to a recent
interaction.
\end{abstract}

\keywords{quasars: absorption lines --- interstellar
medium --- superbubbles}


\section{Introduction}
 
Strong {\MgII} absorbers show complex kinematic structure consistent
with a mixture of radial infall from the halo and a rotating disk
(\cite{Lanzetta92}; \cite{charlton98}; \cite{Churchill01}).  A large
variety of phenomena can contribute to the profiles, such as {\HI}
regions, {\HII} regions, high--velocity halo clouds, and superbubbles.
Strong {\MgII} absorbers almost always have $L_B > 0.08~L_B^*$
galaxies within $35$ kpc (\cite{sdp94}; \cite{steidel95}).  One
particular absorption system, at $z=0.7443$, in the spectrum of
MC~$1331+170$ shows multiple {\it pairs\/} of lines, each with $\Delta
v \sim 30$~{\kms}. There are four strong optical sources within
$5\arcsec$ of the quasar, one of which is an edge--on spiral.  These
are the only detected optical sources which could have impact
parameters less than $35$~kpc.

OB associations are capable of producing large shells of swept up ISM
using the energy input of stellar winds and multiple supernovae.  Such
structures, known as superbubbles, have been observed in nearby
galaxies ({\cite{kamphuis}}; \cite{Sungeon99}) as well as in our own galaxy
(\cite{heiles79}; \cite{McClure00}).  {\MgII} and {\FeII}
absorption would arise within the cold superbubble shell, appearing as
pairs of lines from the oppositely expanding sides of the shell
intersecting the line of sight.

In our galaxy, superbubbles are detected in {\HI} as filaments, with
inferred radii ranging from $100$~pc to $2$~kpc (e.g.\
\cite{heiles79}).  Extragalactic superbubbles are manifest as holes in
the {\HI} map of face--on galaxies or as {\HI} emission--line
filaments in edge--on galaxies.  Single structures of this type are
observed with sizes similar to Galactic bubbles and masses between
$10^7$ and $10^8$~$M_{\odot}$.  Smaller shells (even single--star
stellar wind bubbles) are also thought to exist, but cannot be
detected in external galaxies.  Superbubbles larger than $1$~kpc in
radius are thought to be formed by either a merging of separate shells
or by a process of self--propagating star formation (e.g.\
\cite{bomans96}).  The number of bubbles in a given galaxy varies
dramatically from almost none to several hundred, perhaps depending on
the galaxy's history of interactions.

Stellar wind bubbles for a single early--type star were first modeled
by Weaver {\etal}(1977\nocite{weaver77}).  Mac Low \& McCray
(1988\nocite{macLow88}) extended this concept to clusters of O and B
stars, proposing Galactic superbubbles as the explanation for the
{\HI} filaments observed by Heiles (1979\nocite{heiles79}).  According
to these models, a superbubble begins as an oversized stellar wind
bubble in the ISM around an OB association.  The swept up material
from this cavity quickly collapses into a cold, dense shell with a
density greater than four times that of the ambient ISM. The shell
begins expanding with the onset of the first supernovae in the OB
association, which contribute their energy and metals to the
superbubble.  During this, the adiabatic phase, the supernovae can be
approximated as a continuous energy source and the bubble radius
increases as $t^{3/5}$.  The interior is sparse, highly ionized, and
isobaric during this phase. The continuing energy input maintains the
expansion at a rate of $t^{1/2}$, in excess of the rate expected for a
pure snowplow solution (\cite{macLow88}).

In this paper, we explore the hypothesis that the line splittings in
the $z=0.7443$ absorber arise from expanding superbubbles in the
vicinity of an observed edge--on spiral galaxy (\cite{Churchill95}).
In \S~\ref{sec:data}, we present the HIRES/Keck absorption profiles of
detected low ionization transitions and an archival {\it HST\/}/WFPC2
image of the quasar field.  In \S~\ref{sec:field}, we constrain the
redshifts and luminosities of the four galaxies in the quasar field
using the galaxies' colors and magnitudes.  We consider the
possibility that galaxy pairs and compact groups give rise to the
observed {\MgII} absorption in \S~\ref{sec:kinmod}.  In
\S~\ref{sec:simulations}, we provide a description of the
photoionization models used in testing the superbubble hypothesis.  We
discuss our results and summarize in
\S~\ref{sec:discussion}.

\section{Data and Analysis}
\label{sec:data}

\subsection{HIRES/Keck Spectra}
\label{sec:spectra}

The spectrum of the $z=2.084$, $V=16.7$ quasar MC~$1331+170$ was
obtained with the W. M. Keck Telescope's HIRES/Keck spectrograph
(\cite{Vogt94}) on the night of 19 March 1993.  The slit width was
0.861{\arcsec} (yielding $R=45,000$).  A $2 \times 2$ pixel binning
(in both the cross--dispersion and dispersion directions) was used
during read out.  The resulting resolution is $R=34,000$
($v=8.8$~{\kms} per resolution element). The wavelength coverage is
4263 to 6733~{\AA}, with gaps above 5100~{\AA}, because the CCD did not
capture the free spectral range of the echelle orders above this
wavelength.  The spectrum was reduced in the usual manner using
IRAF\footnote{IRAF is distributed by the National Optical Astronomy
Observatories, which are operated by AURA, Inc., under contract to the
NSF.} v2.10.  All wavelengths are vacuum and heliocentric corrected.

There are four {\MgII} systems in the spectrum of MC~$1331+170$ at
$z=0.7443$, $z=1.3280$, $z=1.776$, and $z=1.786$.  Due to its unusual
structure, our focus is on the $z=0.7443$ system.  The detected
transitions ({\FeII}~$\lambda 2587$, $\lambda 2600$, {\MgIIdblt},
{\MgI}~$\lambda 2853$) for the $z=0.7443$ system are shown in
Figure~\ref{fig:spectrum}.  The profile fits (solid curves) were
obtained using Voigt profile decomposition.  This yields the number of
components, their line--of--sight velocities, column densities, and
Doppler $b$ parameters.  We used MINFIT, a $\chi^2$ minimization code
of our own design (\cite{thesis}).  The ticks above the normalized
continua in Figure~\ref{fig:spectrum} give component velocity
centroids.  Table~\ref{tab:vptab} lists the resulting parameters from
the Voigt profile fitting.

The equivalent width of the $z=0.7443$ system is
$W_r(2796)=1.81$~{\AA}.  An {\it HST\/}/STIS spectrum that covers the
{\CIVdblt} doublet is available in the archive (Program 7271).
Unfortunately, the signal to noise is $\sim 1$ at the position of the
{\CIV} absorption, yielding no useful equivalent width limit.  The
other systems are also strong\footnote{The {\MgII} at $z=1.3284$ is
blended with the {\FeII} from the higher redshift systems (Steidel \&
Sargent 1992) and a precise equivalent width measurement is not
possible, but we estimate $W_r(2796)=0.8$~{\AA}.}.  Damped {\Lya}
absorption (DLA), $N({\HI})\ge 2\times 10^{20}$~{\cmsq} is observed in
the $z=1.776$ system (\cite{Chaffee88}).

Rao and Turnshek (2000\nocite{Rao00}) found that 50\% of {\MgII}
absorbers with $W_r(2796)>0.5$~{\AA} and $W_r(2600)>0.5$~{\AA} are
DLAs (also see \cite{Boisse98}).  This criterion is satisfied by the
$z=0.7443$ system, but the {\Lya} of this system is below the Lyman
limit break of the $z=1.776$ DLA and cannot be measured.  Neutral
hydrogen 21--cm measurements of the $z=0.7443$ system (\cite{Lane00})
give $N({\HI})<1.5 {\times} 10^{18} \left< T_s \right>$~{\cmsq}, where
$\left< T_s \right>$ is the spin temperature.  The minimum spin
temperatures of DLAs ($100-200$~K) occur when they lie in spiral
galaxies (e.g.\ \cite{Kanekar}).  Even if we are looking through the
disk of a spiral galaxy (as we suspect), the 21--cm data do not
restrict the nature of the absorber.  However, the {\MgII} absorption
arising from known DLAs has the characteristic of being totally
saturated across the entire profile (\cite{archive2}) and typically
exhibits a total velocity spread of $50-100$~{\kms}.  These features
are not present in this system, suggesting that it is not a DLA.

\subsection{HST/PC2 Image}
\label{sec:image}

Images of the MC~$1331+170$ field obtained through the F702W ($R$) and
F814W ($I$) filters were retrieved from the archive (Program 5351,
{\cite{Lebrun}}).  Pipeline--reduced images for each filter were
averaged and cleaned using the IRAF task GCOMBINE.  The task
COSMICRAYS was used to remove hot pixels.  To increase the
signal--to--noise ratio for presentation purposes, the $R$ and $I$
band images were combined.  For our analysis, we use the photometry as
presented in Section~$2$, Table~$8$ of Le Brun \etal
1997\nocite{Lebrun}.  Two realizations of this {\it HST\/}/WFPC2
image, with the quasar centered in the PC, are presented in
Figure~\ref{fig:field}.  The left--hand panel has a larger dynamic
range and shows three resolved objects within $5\arcsec$ of the
quasar, labeled in order of angular distance from the quasar.  An
additional source at $\simeq 0.8\arcsec$ is apparent upon
point--spread function (PSF) subtraction of the quasar (Figure~16, of
{\cite{Lebrun}}).  In the right--hand panel, G$5$ is emphasized.

\section{The MC~$1331+170$ Field}
\label{sec:field}

\subsection{Procedure}
\label{sec:procedure}

The photometry of Lebrun \etal(1997\nocite{Lebrun}) gives a limiting
magnitude of $26$ in $R$, which implies that unresolved sources at
$z=0.7443$ brighter than $\sim 0.04~L_B^*$ should be seen in the
image.  Because strong {\MgII} absorbers nearly always have $L_B >
0.08~L_B^*$ counterparts within $35$~kpc (\cite{sdp94};
\cite{steidel95}), we expect to detect the galaxy giving rise to the
$z=0.7443$ absorption.

There are four galaxies which could be within $35$~kpc of the quasar
in the {\it HST\/}/WFPC2 image (Figure~\ref{fig:field}, G$2$--G$5$,
numbered as in \cite{Lebrun}).  Inspection of the sources shows an
edge--on spiral morphology for G$5$ and an elliptical shape for G$3$
(also see \cite{Lebrun}).  G$4$ shows some structure (possibly
irregular) and the location of G$2$ in front of the quasar makes
morphology difficult to determine.

Unfortunately, redshifts for G$2$--G$5$ have not been obtained.  We
must, therefore, determine how likely it is that each candidate galaxy
is associated with the absorber at each of the redshifts.  This will
allow us to assess which galaxy/galaxies could give rise to the
$z=0.7443$ absorption.  We do this by comparing the apparent
magnitudes and colors of each galaxy to redshifted model spectral
energy distributions (SED) for different morphological types.
However, we appreciate that there may not be a one--to--one
correspondence between absorbers and galaxies. Two or more of the
observed galaxies could give rise to absorption and lie at the same
redshift (\cite{Bowen95}).  Additionally, DLA absorption can arise
from relatively low luminosity and/or low surface brightness galaxies
which are not seen in the image (e.g.,
\cite{Rao00}; \cite{Bouche00}; also see \cite{Bowen2000}).

In order to obtain the $L_B$ values expected for each galaxy at each
redshift, we first obtained $B$ and $R$ magnitudes from model SEDs for
a set of standard morphologies (\cite{Bruzual93};
\cite{Kinney96}). Models for E, S0, Sab, Sbc, and Scd galaxies
($\mu$ $=0.95, 0.30, 0.20, 0.10, 0.01$, respectively\footnote{$\mu$ is
the fraction of the mass in the stars after one Gyr (Bruzual 1983)})
were modelled with a $16$~Gyr stellar population with exponentially
decreasing star formation.  A redder elliptical galaxy (E2) was
modeled using a composite spectrum of observed ellipticals from Kinney
\etal 1996\nocite{Kinney96}.  For an irregular galaxy (Im), the observed
SED of NGC 4449 was adopted (\cite{Bruzual93}).  Two starburst models
were also considered, both with a constant star formation rate,
assuming a Salpeter IMF.  The ages of the starburst galaxies
(Starburst 1 and Starburst 2, Bruzual \& Charlaut
1993\nocite{Bruzual93}) are $10^6$ and $10^7$ years.  

The K--corrections were performed by redshifting the model SEDs to the
absorber redshifts and convolving them with the F702W and F814W filter
response curves.  A zero--point normalization was performed using the
measured $R$ band apparent magnitudes listed in
Table~\ref{tab:galanal} (also see Table~8 in \cite{Lebrun}).  Finally, the
$M_B$ values were calculated using $H_0=50$~{\kmsM} and $q_0=0.05$.
Figure~\ref{fig:colors} shows $R-I$ color vs.\ redshift for the
standard morphological types.  The dotted lines on the figure
correspond to the redshifts of the {\MgII} absorptions systems in the
MC~$1331+170$ spectrum and the dashed lines to the colors of the four
galaxies, G$2$--G$5$.  Intersections of the lines represent permitted
locations on the diagram of the galaxies in the field and allow us to
constrain their type at each possible redshift.

\subsection{Results}
\label{sec:results}

Table~\ref{tab:galanal} summarizes the permitted morphological types
and their calculated $L_B$ (in units of $L_B^*$, using $M_B^*=-20.9$)
at each absorber redshift (excepting $z=1.786$ because of its
proximity to $z=1.776$).
  
The errors on the color of G$2$ and magnitude might be quite large due
to the PSF subtraction of the quasar.  The color of G$2$ makes it
unlikely to be at $z=0.7443$, and it must be an elliptical if it is at
$z=1.3284$ or $z \sim 1.8$.  If only its color is considered, it could
be an E or S0 at $z \sim 1.8$, but the derived $M_B$ values give $L_B
\sim 15~L_B^*$ for an E and $L_B \sim 5~L_B^*$ for an S0.  Such
luminous galaxies are exceedingly rare and we conclude that G$2$ is
probably at $z=1.3284$.

The morphology and color of G$3$ seem to indicate that it is an
early--type galaxy, but its redshift is difficult to constrain.  If at
$z=0.7443$, G$3$ would have $L_B \sim 0.1~L_B^*$ and could be an E,
S0, or early--type spiral.  Due to the low luminosity, it is
uncertain whether G$3$ would be giving rise to strong {\MgII} absorption
at this redshift.  G$3$ could be an S0 at either $z=1.3284$ with $L_B
\sim 1~L_B^*$ or $z \sim 1.8$ with $L_B \sim 4.3~L_B^*$.  If it is the 
latter, the galaxy would be unusually luminous.  Therefore, it is most likely
an E or S0 at $z=0.7443$ or $z=1.3284$.

G$4$ is very blue and would have to be actively star--forming to be at
$z=0.7443$.  Just as with G$3$, however, it may not be luminous enough
($L_B \sim 0.1~L_B^*$) at this redshift to be expected to give rise to
strong {\MgII} absorption.  The irregular structure observed in the
image may indicate an Im galaxy.  The color and luminosity ($L_B
\sim 0.1~L_B^*$) are also consistent with such a morphology at either
$z=1.3284$ or $z \sim 1.8$.  The color--redshift curves converge
toward $R-I \sim 0$ at high redshift for the spiral types, allowing an
Scd or Sbc identification for G$4$ at $z \sim 1.8$.

The most striking galaxy in the MC~$1331+170$ field is G$5$, an
edge--on spiral.  This is our favored candidate for a $z=0.7443$
absorber, with a luminosity of $L_B \sim 1~L_B^*$.  The low
luminosities of G$3$ and G$4$ make them marginal cases for producing
such strong {\MgII} absorption.  Additionally, the luminosity of G$5$
becomes larger for the higher absorber redshifts ($\sim 2-3~L_B^*$);
thus, it $\sim 10$ times less likely to be at the higher redshifts.
The color of G$5$, however, does not constrain its redshift and allows
it to be a star--forming galaxy at any of the higher redshifts.  If
this is so, G$3$ or G$4$ would be responsible for the $z=0.7443$
system, but their small luminosities at this redshift make this
unlikely.

In conclusion, G$5$ is the galaxy most likely to be producing
$z=0.7443$ absorption, but G$3$ and G$4$ could also be contributing to
the absorption at that redshift.  If so, they would have small
luminosities that make them border--line candidates for strong {\MgII}
absorption.

\section{Kinematic Model Simulations of Groups and Pairs}
\label{sec:kinmod}

Though G$5$ is the best candidate for the $z=0.7443$ absorber, G$3$
and G$4$ also have colors and luminosities consistent with
this redshift.  Furthermore, the blue colors of G$4$ and G$5$ seem to
indicate recent star formation, which could easily have been triggered
by an interaction between two or more galaxies.  As such, two
compelling possibilities are that the line of sight passes through
several members of a compact group of galaxies or through a galaxy pair,
giving rise to the observed {\MgII} absorption profile.

Charlton and Churchill (1998)\nocite{charlton98} found that the
kinematics of strong {\MgII} absorbers can be described by assuming
that the absorbing clouds are located both in the disks and halos of
normal galaxies.  In order to synthesize an absorption line profile, we run a
line of sight through a model galaxy with a given orientation, impact
parameter, and size (where the size of the galaxy is defined by its
luminosity).  We used the luminosity and impact parameter from
Table~\ref{tab:galanal} as the input for each galaxy model.  The
orientation of galaxy G$5$ was set to be edge--on but, when the other
galaxies were used, their orientations were chosen randomly. Each
galaxy model was chosen from $75$D/$25$H Hybrid$2$ in Charlton \&
Churchill (1998)\nocite{charlton98}, with $75$\% disk clouds and
$25$\% halo clouds, by number.  In order to simulate absorption from
either compact groups or galaxy pairs, the individual synthetic
absorption profiles were shifted in velocity based on observed compact
group or galaxy pair kinematics.  We generated $1000$
realizations of synthetic spectra for each.

\subsection{Compact Groups}
\label{sec:groups}

The median velocity dispersion of compact groups, $200$~{\kms}
(\cite{hickson92}), results in profiles that have velocity spreads
larger than in the $z=0.7443$ system.  A smaller velocity dispersion
was required so that our models did not produce synthetic profiles
with total velocity spreads inconsistent with the data.  Therefore,
we chose the galaxy relative velocities from HGC 22, the compact group
with the minimum velocity spread ($100$~{\kms}, \cite{hickson92}).
Thus, we have already restricted ourselves to a limited region of
parameter space; this is not a ``typical'' compact group.

Twenty--five randomly selected synthetic spectra from our
three--galaxy compact group models (using G$3$, G$4$, and G$5$) are
shown in the upper panel of Figure~\ref{fig:kinmods}.  Clearly, the
velocity spreads of the clouds in the simulated spectra tend to be
greater than that in the $z=0.7443$ system, even with the bias towards
a small dispersion in galactic velocities.  Also, the saturation
levels in the simulated spectra tend to be higher, implying that our
particular line of sight passes through less material than in an
average model compact group.  Finally, the regular structure of pairs
in absorption, separated by $\sim 30$~{\kms}, is not a characteristic
feature of lines of sight through model compact groups.  

\subsection{Galaxy Pairs}
\label{sec:pairs}

In order to reduce the total velocity spread and saturation, we also
considered models of galaxy pairs, assuming that either G$3$ or G$4$
is at $z=0.7443$ along with G$5$.  Again, G$5$ was taken to have an
edge--on orientation, and the other galaxy was taken to be randomly
oriented.  The velocity difference between the galaxy pair is tuned to
$\sim50$~{\kms}.  In fact, this is a relatively typical value for
field galaxy pairs.

Twenty--five randomly selected galaxy pair synthetic spectra are
displayed in the bottom panel of Figure~\ref{fig:kinmods}.  Compared
to the compact group synthetic profiles, a larger percentage of the pair
profiles have velocity spreads and saturation levels consistent with
the observed {\MgII} profile.  However, the regular absorption line
pairs we observe in the spectrum of MC~$1331+170$ are only rarely
produced, even in this simple galaxy pair model.

\subsection{Kinematic Model Results}
\label{sec:kinresults}

The $z=0.7443$ absorber toward MC~$1331+170$ was selected for this
study because of its uniquely regular properties.  As such, this is a
biased study.  Nonetheless, we have seen that, in $1000$ Monte--Carlo
simulations of galaxy pairs and in $1000$ simulations of compact
groups, we can only rarely produce a profile with such low levels of
saturation and regular velocity splittings.  However, it is difficult
to quantify the shapes of the profiles in such a way that we can make
direct statistical comparisons.  We remind the reader that we used a
compact group with an unusually small velocity dispersion.  Therefore,
based on this and on a visual inspection, we conclude that the
observed profile is unlikely to arise from either a compact group or
galaxy pair environment.

\section{Superbubble Simulations}
\label{sec:simulations}

The spectrum of the $z=0.7443$ absorption system appears to be
kinematically consistent with a line of sight intersecting the
oppositely expanding sides of the ``cold'' shells of multiple
superbubbles (\cite{Churchill95}).  In order to test this hypothesis,
we calculated what the column densities of {\MgII} and {\FeII} should
be in the context of such a model.

\subsection{The Model}
\label{sec:simmod}

The global properties of our model superbubbles are governed by the
similarity solutions of Mac Low \& McCray (1988)\nocite{macLow88}
(also see \cite{weaver77}).  Models of the cold shell of the
superbubbles were performed using Cloudy, version $94.00$
(\cite{ferland96}), taking into account chemical and ionization
conditions.  The photoionization source consists of the ultraviolet
background from quasars (\cite{Haardt}) and the O and B stars driving
the expansion of the superbubble. The intensity of the Haardt--Madau
background is set to a constant hydrogen--ionizing photon number
density of $10^{-5.2}$~{\cc} (as appropriate for $z \sim 1$).  The
stellar contribution depends on the luminosity of the parent OB
association and the radius of the superbubble.  Based upon our models,
we find that for luminosities greater than $\sim
10^{38}$~ergs~s$^{-1}$, the stellar spectrum dominates at ages less
than $\sim 10^7$~years.  Most of the models, however, are dominated by
the Haardt--Madau spectrum.

We assume the ISM to have a constant initial atomic density of
$n_0$~{\cc}.  An OB association residing in this medium generates a
total luminosity of $L_{38}$ (in $10^{38}$ ergs~{\cc}) through a
series of supernova explosions.  If each supernova produces $\sim
10^{51}$~ergs and the OB association lasts $50$ million years, the
time--averaged luminosity is related to the number of supernovae by
$N_{SN} \simeq 580~L_{38}$.  The atomic density inside of the swept up
shell is $n_e$ (expressed in units of {\cc}) and the
metallicity is $Z$, in solar units.  The energy input from
supernovae drives the bubble radially outward into the ISM.  The
similarity solution (\cite{macLow88}) gives the radius of the bubble,
\begin{equation}
R = 0.27~\left(n_0/L_{38}\right)^{-0.2}~t_7^{0.6} \quad{\rm kpc},
\label{eq:radius}
\end{equation}
where $t_7$ is the age in units of $10^7$~years.  The shell expansion
speed then follows as
\begin{equation}
\dot R = 16~\left(n_0/L_{38}\right)^{-0.2}~t_7^{-0.4} \quad{\kms},
\label{eq:speed}
\end{equation}
which is constrained by the observed velocity splitting, $\Delta v$,
of the absorption lines according to
\begin{equation}
\dot R = 0.5~\Delta v~\left(1 - z^2/R^2\right)^{-1/2},
\label{eq:splitting}
\end{equation}
where $z$ is the impact parameter (to the center of the bubble) of the
line of sight.  This equation is not very sensitive to impact
parameter unless $z/R \sim 1$.  From the observed profile, $\Delta v
\sim 30$~{\kms}; thus, we set $\dot R = 15$~{\kms}.

If we choose $L_{38}$ and $t_7$, we can calculate $R$ and $N_{tot}$
for input to the Cloudy model, where $N_{tot}$ is the total hydrogen
column density.  The radius is found by combining equations
(\ref{eq:radius}) and (\ref{eq:speed}), giving
\begin{equation}
R = 0.017~\dot R~t_7 \quad{\rm kpc}.
\label{eq:radius2}
\end{equation}
From equation (\ref{eq:radius}), $n_0$ is given by
\begin{equation}
n_0 = L_{38}~\left(16~t_7^{-0.4}/\dot R\right)^5 \quad{\cc}.
\label{eq:ismdens}
\end{equation}
Finally, $N_{tot}$ is
\begin{equation}
N_{tot} = 10^{21}~R~n_0 \quad{\cmsq}.
\label{eq:column}
\end{equation}

The luminosity, $L_{38}$ ($0.01$,$0.1$,$1$,$10$), was chosen for OB
associations containing between $5$ and $5,000$ stars (e.g.\
\cite{McKee97}; \cite{Bresolin98}).  The age, $t_7$ ($0.5$,$1$,$2$,$4$),
was chosen to represent different points in the evolution of a bubble
expanding for $50$ million years (\cite{macLow88}).  The Cloudy models
of the shell required that we also input $\log n_e$ ($-2.5$ to $2$,
with intervals of $0.25$).  Cloudy was then run for every combination
of the chosen parameters ($L_{38}$, $t_7$, $n_e$, and $Z$).  The
metallicity range ($\log Z \la 0$) was based upon observational
constraints (see \S~\ref{sec:simresults}).  Each simulation was halted
when the hydrogen column density ($N_{tot}$) calculated in Equation
\ref{eq:column} was reached.

\subsection{Simulation Results}
\label{sec:simresults}

For each $L_{38}$ and $Z$, Cloudy grids were generated. In
Figure~{\ref{fig:grids}}, we show a plot of output grids from Cloudy
photoionization models for the $\log Z=-1$ and $\log Z=-0.5$ cases.  The
thickness of the plane parallel supernova shell is plotted versus its
{\MgII} column density.  Each panel represents a different luminosity
of the parent OB association (given by $L_{38}$).  The solid lines
represent the ages of the model superbubbles, corresponding (from left
to right) to $t_7=4$, $2$, $1$, and $0.5$.  The dashed lines represent
the hydrogen number density of the shell, which ranges from (top to bottom)
$\log n_e=-3$ to $\log n_e=2$~{\cc} in logarithmic intervals of $0.25$.  We
picked the regions of parameter space which are in the observed {\MgII}
column density range (see Table~\ref{tab:vptab}) and are consistent
with observed superbubbles (\cite{heiles79}).  These regions are
enclosed by the dotted boxes.

The models produce values for $n_e / n_0$ over a large range, the
maximum being $\sim 1000$.  We only plot models for which the shell is
more dense than the ISM ($n_e > n_0$, where $n_0$ is found from
Equation {\ref{eq:ismdens}}).  This will probably be the case unless
there is an extreme ISM density gradient (\cite{macLow89}) or we are
passing through a cloudy medium (\cite{Silich}).  Real superbubbles
have shell densities greater than four times the ISM densities,
perhaps greater than ten times if the shells are overstable
({\cite{ryu88}}).

It may seem counter--intuitive that {\MgII} column density decreases
as time increases, but the successive $t_7$ grid lines do not
represent an evolutionary sequence.  The decrease is a consequence of
holding shell speed constant for all models. In order to produce the
same speed at a later time, the ISM density muchbe lower
(providing less resistance).  Thus, the different $t_7$ lines
represent varying initial conditions (see Equation~\ref{eq:speed}).
If you have a smaller ISM density, then you will have a smaller amount
of material swept up into the shell, resulting in a smaller {\MgII}
column density.

Increasing the luminosity of the clusters corresponds to a shift of
the grid toward higher {\MgII} column densities and thicker shells.
Increasing metallicity causes a shift in the grid towards lower
{\MgII} column densities.  We have set constraints on the value of
$L_{38}$, so we can now constrain $Z$ to fit observational parameter
space (the dotted box).  The three left--hand panels of
Figure~\ref{fig:grids} show $\log Z=-1$, which is consistent with the
observed {\MgII} column densities for $L_{38}=0.01$ or $L_{38} = 0.1$.
A grid with an increased metallicity would require a lower $L_{38}$
(as apparent in the three right--hand panels).  If $\log Z>0$, the models
require a luminosity that is too low for an OB association ($L_{38} <
0.01$).  The largest OB associations contain $\sim 7000$ supernova
progenitors ({\cite{McKee97}}), corresponding to $L_{38} \sim 10$ (not
shown in Figure~\ref{fig:grids}).  Even under these extreme conditions, 
we cannot place a useful lower limit on the metallicity in the shell.

We can not obtain general lower and upper limits on $t_7$ (though
$t_7<5$ is a physical limit in our models, see
\S~\ref{sec:simmod}). The ages are limited, however, for particular
combinations of $L_{38}$ and $Z$.  For example, the $L_{38} = 0.1$,
$\log Z = -1$ grid in Figure~{\ref{fig:grids}} has only $1 < t_7 < 4$
lying within the allowed parameter space (dotted box).  Presumably,
the multiple bubbles giving rise to our spectrum could arise from the
same burst of star formation and have similar ages and metallicities.
For this to be true, a logarithmic spread of $1.5$ is required in
$L_{38}$ in order to reproduce the entire range of {\MgII} column
densities.

As with $t_7$, we cannot obtain general lower and upper limits on the
shell density, $n_e$, but we can constrain it for particular
combinations of $L_{38}$ and $Z$.  As can be seen in
Figure~{\ref{fig:grids}}, increasing $L_{38}$ raises the grid and
brings higher shell densities into the allowed parameter space.  For
example, the $L_{38} = 0.1$, $\log Z = -1$ grid has only $-1.5 > \log
n_e > 1$~{\cc} lying within the allowed parameter space.

\section{Conclusion and Discussion}
\label{sec:discussion}

The line pairs in the $z=0.7443$ {\MgII} absorption system in the
spectrum of MC~$1331+170$ are strongly suggestive of shells of
expanding superbubbles.  The identity of the host of the $z=0.7443$
absorption system is not immediately obvious, but is likely to be the
spiral galaxy (G$5$) seen in the {\it HST}/WFPC2 image
(Figure~{\ref{fig:field}}).  G$3$ and G$4$ are also candidates for
being at $z=0.7443$ and may form a compact group or pair with G$5$.
However, kinematic models of galaxy pairs and compact groups seem to
yield velocity spreads larger than the observed $\sim 200$ {\kms} and
rarely reproduce the symmetry seen in the MC~$1331+170$ system.  This
evidence seems to suggest that the absorption arises in a single
galaxy, either from superbubbles or random cloud distributions.
Photoionization models, considering the conditions of the parent OB
association and the surrounding medium, support the superbubble
hypothesis without requiring fine--tuning of the specific conditions
from which the superbubbles arise.

The models of superbubbles do, in fact, produce the observed {\MgII}
column densities and velocity splittings of the $z=0.7443$ system, but
there are still some discrepencies.  The $\sim 250$~{\kms} spread of
our profile is the most problematic feature in the context of disk
superbubbles.  A typical $L^*$ galaxy has a rotation velocity ($V_0$)
of $\sim 220$~{\kms}.  If one assumes that the superbubbles follow the
disk rotation, the maximum spread ($V_{max}$) over which they could
occur is $V_{max} = V_0 (1 - D/R_g)$, where $D$ is the impact
parameter and $R_g$ is the maximum galactic radius at which
superbubbles can occur.  The parameter $R_g$ is likely to be
approximately equal to the optical radius of the galaxy.  In M$101$,
superbubbles are seen out to $30$~kpc (\cite{kamphuis}).  If we
assume that $R_g$ can be as large as $\sim 40$~kpc, we get only
$V_{max} \sim 100$~{\kms}.  For G$5$ at $z=0.7443$, the impact
parameter is $20\h^{-1}$~kpc.

There are several possible explanations for the large spread in the
$z=0.7443$ {\MgII} absorber: 1) A merger is inducing globular cluster
formation in the halo (\cite{Ashman}; \cite{Zepf}) and the clusters
are creating superbubbles.  2) Superbubbles are forming within tidal
debris from a recent interaction (\cite{Kniermann00};
\cite{Gallagher01}).  A line of sight through tidal debris can create
velocity spreads as large as $1000$~{\kms} (\cite{Gallagher01}).  3)
Some of the superbubbles are outliers in velocity space, possibly
located within a companion dwarf galaxy (\cite{macLow99}).  4) Some of
the absorption lines are not from superbubbles and lie outside of the
disk.  This absorption could arise in halo clouds or in dwarf galaxies.

Womble, Junkkarinen, and Burbidge (1991){\nocite{Womble91}} point out
that large velocity spreads in edge--on galaxies are better explained
by chimneys, worms, or superbubbles blowing out into the halo.  These
are processes known to occur at the end of a superbubble's lifetime
when the shell accelerates through a density gradient into the halo.
The shell then fragments due to Rayleigh--Taylor instabilities and the
bubble begins expelling material out of the disk.  This process is
quite messy, however, and is not likely to give rise to such evenly
split lines.

\begin{figure*}
\figurenum{1}
\plotfiddle{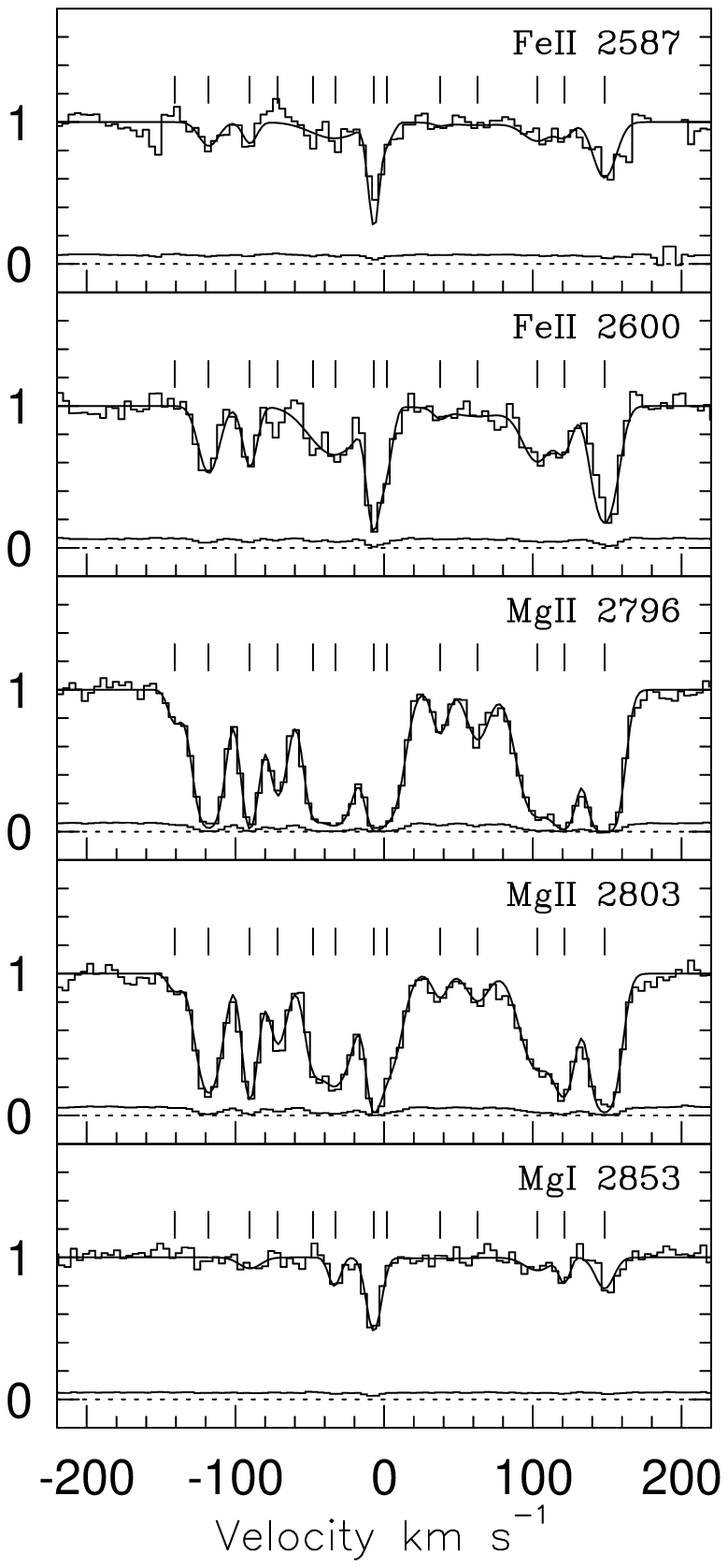}{6.in}{0}{60}{60}{-300}{0}
\caption{
The HIRES/Keck spectrum of the z=$0.7443$ system toward Q1331+170 showing 
transitions detected at $>5\sigma$ aligned in rest frame velocity space.
The solid lines represent a model from a simultaneous Voigt profile fit
to the spectra, with the ticks marking individual components.
}
\label{fig:spectrum}
\end{figure*}

\begin{figure*}
\figurenum{2}
\plotfiddle{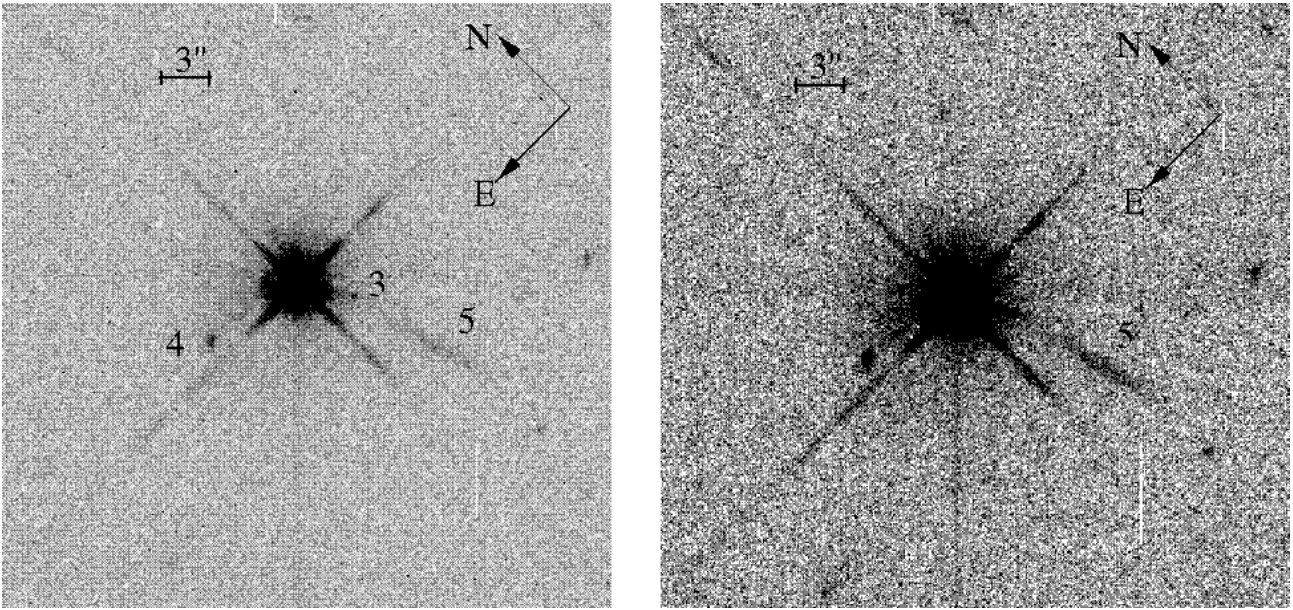}{6.in}{270}{50}{50}{-220}{350}
\caption{
Two images, each with a different dynamic range, of the MC~$1331+170$
field combining the $R$ (F$702$W) and $I$ (F$814$W) filters.  Galaxies
are numbered G$3$ through G$5$, in order of distance from the quasar
(at center).  G$2$ is not labeled because of its proximity to the
quasar and can only be resolved via PSF subtraction.  The left--hand
panel emphasizes G$3$ and G$4$ and the right--hand panel emphasizes
G$5$ (for presentation purposes only).}
\label{fig:field}
\end{figure*}

\begin{figure*}
\figurenum{3}
\plotfiddle{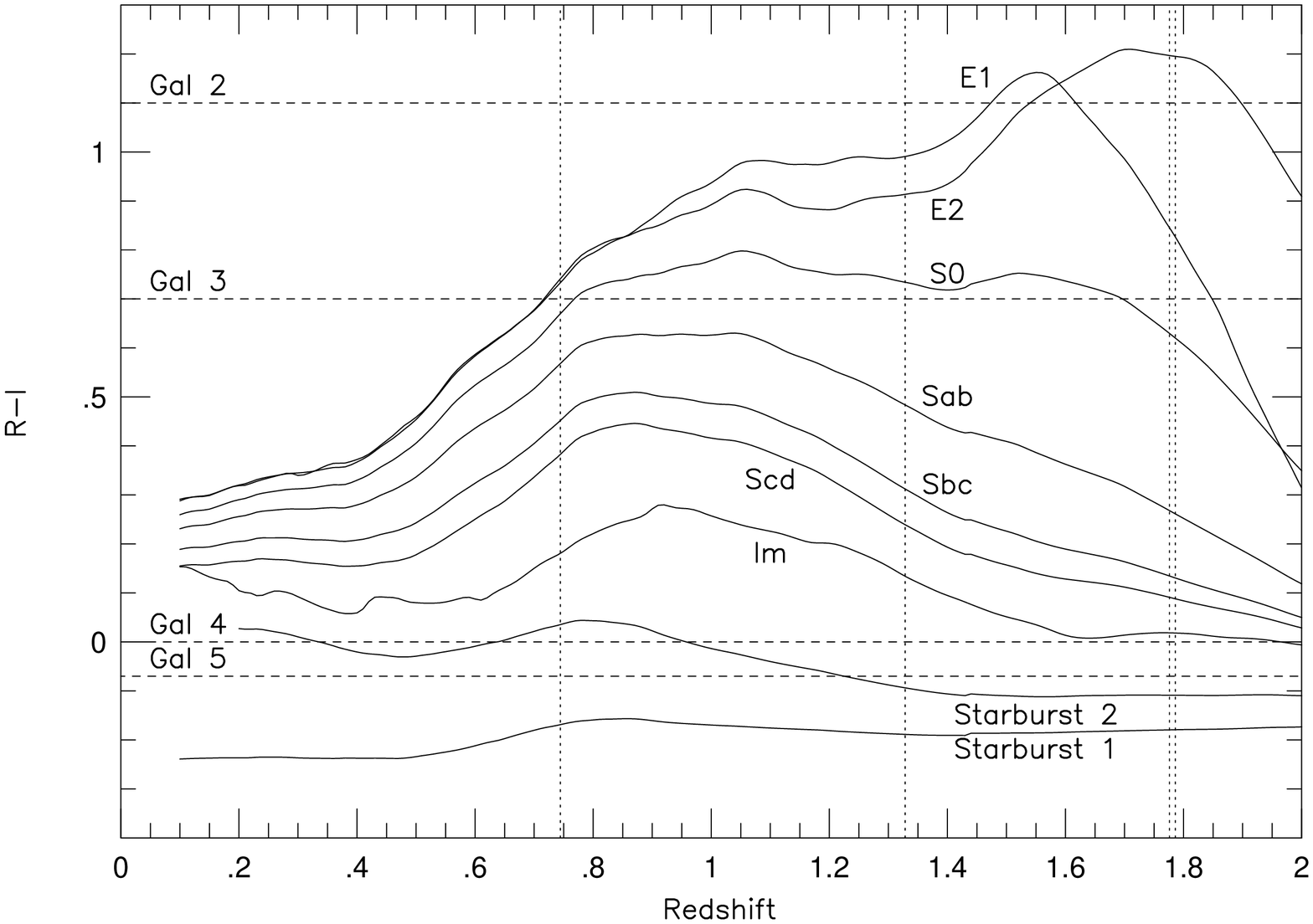}{6.in}{0}{60}{60}{-300}{0}
\caption{
A color-redshift diagram plotted of model spectral energy
distributions representing a variety of morphological types
(\cite{Bruzual93}; \cite{Kinney96}).  The dotted vertical lines
denote the redshifts of the absorption systems in the spectrum of
Q1331+170 and the dashed horizontal lines denote the colors of the
sources in the optical field.  }
\label{fig:colors}
\end{figure*}

\begin{figure*}
\figurenum{4}
\plotfiddle{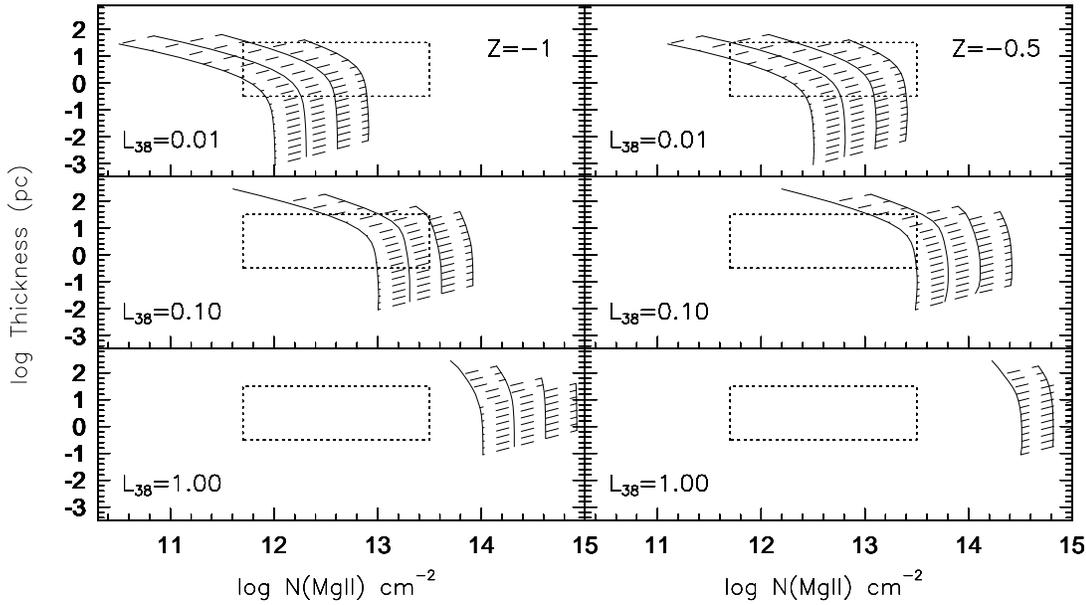}{6.in}{0}{60}{60}{-300}{-100}
\caption{
Output grids from Cloudy photoionization models for the $\log Z=-1$
and $\log Z=-0.5$ cases.  Each plot of shell thickness vs.\ {\MgII}
column density in a column presents a box that represents a different
luminosity input from the parent OB association (given by $L_{38}$).
From left to right, the solid curves represent the ages of the model
superbubbles, corresponding to $t_7=4$, $2$, $1$, and $0.5$.  The
dashed curves represent hydrogen number densities in the shell which
are (from top to bottom) $\log n_e=-3$ to $2$ cm$^{-3}$ in logarithmic
intervals of $0.25$.  The boxes in the center of the plots represent
expected parameter ranges based on observed superbubbles.  }
\label{fig:grids}
\end{figure*}

\begin{figure*}
\figurenum{5}
\plotfiddle{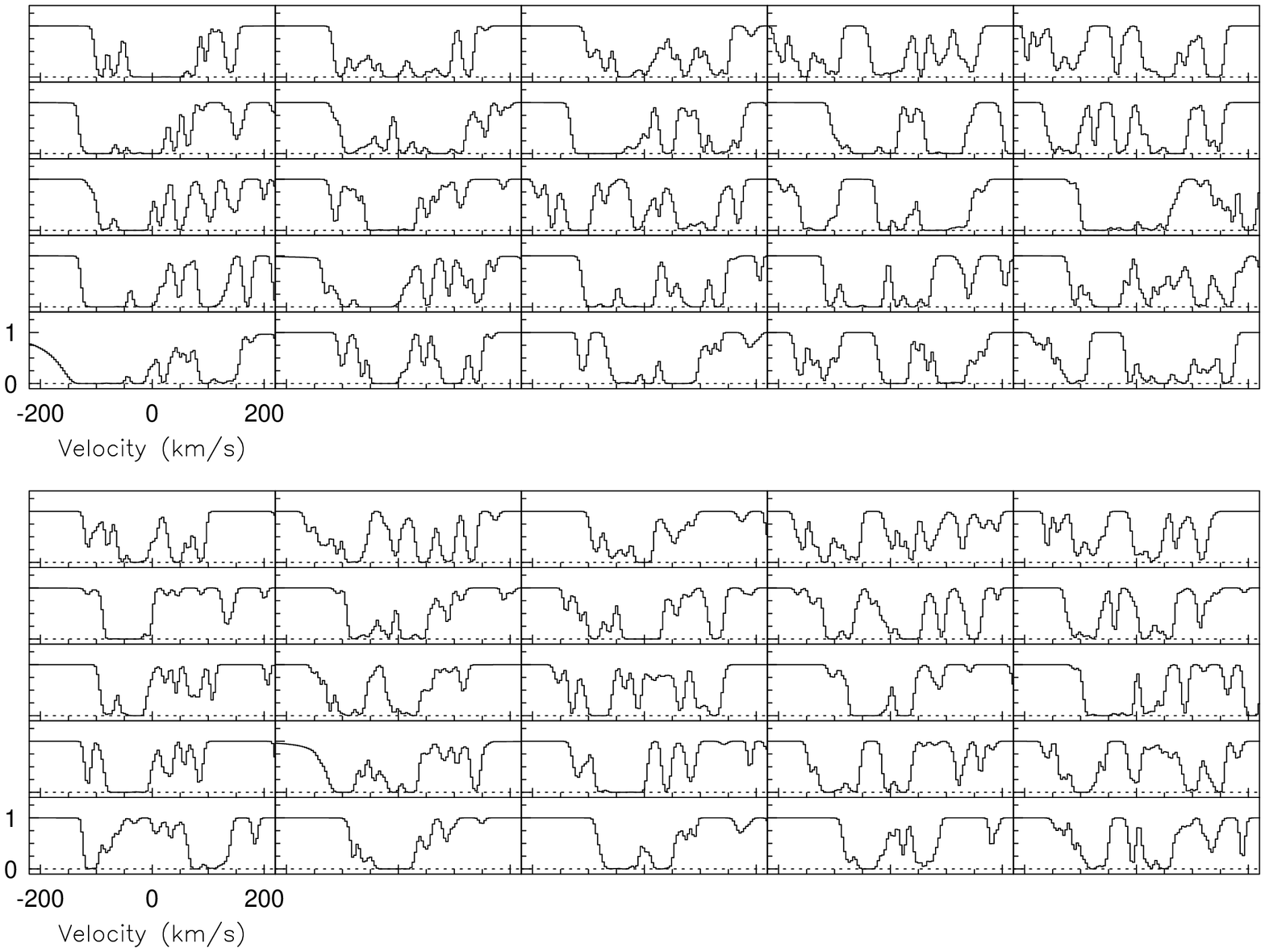}{6.in}{0}{60}{60}{-300}{0}
\caption{
Spectra generated from $25$ models of three--member compact groups (top
panel) and $25$ models of galaxy pairs (bottom panel).  }
\label{fig:kinmods}
\end{figure*}

Our spectrum shows no more than six superbubbles within the line of
sight, each $\sim 10$~pc thick ({\cite{macLow89}}); thus, there must
be a substantial path length through the ISM in between the
superbubbles.  Why don't we see dominating absorption from the ISM
between the superbubbles?  The simplest explanation is that the
filling factor of the superbubbles is so large that the path length
through the ISM makes a negligible contribution to the absorption.
Observations of {\HI} holes in nearby galaxies ({\cite{kamphuis}})
give maximum covering factors of $\sim 0.3$; however, they are only
sensitive down to $R \sim 0.5$~kpc.  Another possible explanation is
that OB associations reside in regions of enhanced density and the
superbubbles sweep the densest material into their shell before
reaching the sparse, intervening material.  Finally, the parent OB
associations themselves are at unusual velocities with respect to most
of the ISM material.  This separation in velocity space may arise if
the bubbles are localized in a certain part of the disk or exist in a
star--forming tidal tail nearby.  The lines at $\sim -100$~{\kms} in
Figure~{\ref{fig:spectrum}} show less symmetry than the others and
could be arising from ISM clouds in the disk.

\acknowledgements
We acknowledge support from the NSF/REU program and NASA/LTSA through
grant NAG5-6399.  We would like to thank Mordecai-Mark Mac Low for his
insightful comments.  We thank Janet Geoffroy for the use of her
spectral integration code which we used to calculate galaxy colors.  We
also benefit from useful conversations with Matt Bershady Niel Brandt,
Robin Ciardullo, Jenn Donley, Rajib Ganguly, Dick McCray, Jane Rigby,
Ken Sembach, and Steinn Sigurdsson.



\input{tab1}      
\input{tab2}      


\end{document}

%% file: tab1.tex

\begingroup
\footnotesize
\begin{deluxetable}{rrrrrrr}
\tablewidth{0pc}
\tablecolumns{7}
\tablecaption{Cloud Properties for Q1331+170}
\tablehead
{ 
\colhead{Cloud} &
\multicolumn{2}{c}{{\MgII}} &
\multicolumn{2}{c}{{\FeII}} &
\multicolumn{2}{c}{{\MgI}} \\
\cline{2-3} \cline {4-5} \cline{6-7}
\colhead{$\left< v \right>$} &
\colhead{$\log N$} &
\colhead{$b$} &
\colhead{$\log N$} &
\colhead{$b$} &
\colhead{$\log N$} &
\colhead{$b$} \\
\colhead{[{\kms}]} &
\colhead{[{\cmsq}]} &
\colhead{[{\kms}]} &
\colhead{[{\cmsq}]} &
\colhead{[{\kms}]} &
\colhead{[{\cmsq}]} &
\colhead{[{\kms}]} 
}
\startdata
  $-140.7$ & $11.81\pm0.08$ & $ 5.01\pm1.68$ & $<11.89$ & \nodata & $<10.80$ & \nodata \nl
  $-117.9$ & $13.15\pm0.01$ & $ 9.09\pm0.16$ & $12.82\pm0.03$ & $ 8.33\pm0.93$ & $<10.78$ & \nodata \nl
  $-90.2$ & $13.05\pm0.01$ & $ 4.70\pm0.12$ & $12.61\pm0.04$ & $ 4.81\pm0.85$ & $11.02\pm1.15$ & $\sim 9.5$ \nl
  $-71.4$ & $12.63\pm0.01$ & $ 7.13\pm0.37$ & $12.07\pm0.19$ & $ 2.19\pm1.57$ & $<10.80$ & \nodata \nl
  $-47.9$ & $12.62\pm0.08$ & $ 6.29\pm0.59$ & $11.62\pm0.06$ & $ 0.31\pm0.08$ & $<10.84$ & \nodata \nl
  $-32.8$ & $13.22\pm0.02$ & $13.22\pm0.68$ & $12.87\pm0.04$ & $13.54\pm1.79$ & $11.21\pm0.22$ & $ 4.03\pm2.87$ \nl
  $ -6.9$ & $13.26\pm0.02$ & $ 4.03\pm0.20$ & $13.25\pm0.04$ & $ 3.57\pm0.26$ & $11.81\pm0.04$ & $ 5.50\pm0.80$ \nl
  $  2.1$ & $13.03\pm0.04$ & $10.33\pm0.58$ & $12.54\pm0.09$ & $ 3.87\pm1.09$ & $<10.75$ & \nodata \nl
  $ 37.6$ & $12.01\pm0.06$ & $ 5.88\pm1.19$ & $<11.82$ & \nodata & $<10.80$ & \nodata \nl
  $ 63.0$ & $12.21\pm0.04$ & $ 8.65\pm1.10$ & $12.52\pm0.22$ & $39.46\pm22.98$ & $<10.80$ & \nodata \nl
  $103.4$ & $13.08\pm0.02$ & $13.33\pm0.56$ & $12.83\pm0.06$ & $12.21\pm1.71$ & $11.21\pm0.28$ & $12.81\pm8.38$ \nl
  $121.4$ & $13.08\pm0.02$ & $ 7.77\pm0.23$ & $12.45\pm0.08$ & $ 6.15\pm1.24$ & $11.09\pm0.30$ & $\sim 2.6$ \nl
  $148.3$ & $13.47\pm0.01$ & $ 9.07\pm0.12$ & $13.31\pm0.01$ & $ 9.75\pm0.38$ & $11.43\pm0.07$ & $ 7.59\pm1.58$ \nl
\multicolumn{7}{c}{ } \nl
\enddata
\label{tab:vptab}
\end{deluxetable}
\endgroup


%% file: tab2.tex

\begingroup
\footnotesize
\begin{deluxetable}{ccccccccccccc}
\tablewidth{0pc}
\tablecolumns{13}
\tablecaption{Q1331+170 Field Analysis Scenarios}
\tablehead
{ 
\colhead{Galaxy} &
\colhead{D} &
\colhead{R} &
\colhead{I} &
\multicolumn{3}{c}{{$0.7443$}} &
\multicolumn{3}{c}{{$1.3284$}} &
\multicolumn{3}{c}{{$1.776$}} \\
\cline{5-7} \cline {8-10} \cline{11-13}
&
&
&
&
\colhead{D} &
\colhead{Type} &
\colhead{Luminosity} &
\colhead{D} &
\colhead{Type} &
\colhead{Luminosity} &
\colhead{D} &
\colhead{Type} &
\colhead{Luminosity} \\
&
\colhead{[\arcsec]} &
&
&
\colhead{[kpc]} &
&
\colhead{[{$L^*$}]} &
\colhead{[kpc]} &
&
\colhead{[{$L^*$}]} &
\colhead{[kpc]} &
&
\colhead{[{$L^*$}]} 
}
\startdata
    $2$ & $0.75$ & $24.9$ & $23.8$ & $3.61$ & \nodata & \nodata & $4.36$ & $E$  & $3.57$ & $4.63$ & $E$  & $14.5$ \nl
        &        &        &        &        &         &         &        &      &        &        & $S0$ & $5.14$ \nl
    $3$ & $1.58$ & $25.1$ & $24.4$ & $7.61$ & $S0$    & $0.08$  & $9.18$ & $S0$ & $1.09$ & $9.76$ & $S0$ & $4.28$ \nl
        &        &        &        &        & $E$     & $0.08$  &        &      &        &        &      &        \nl
    $4$ & $2.86$ & $24.2$ & $24.2$ & $13.77$ &   $SB$ & $0.08$  & $16.62$ & $SB$ & $0.163$ & $17.67$ & $SB$ & $0.208$ \nl
        &        &        &        &         &   $Im$ & $0.12$  &         & $Im$ & $0.498$ &         & $Im$ & $0.822$ \nl
    $5$ & $3.86$  & $21.40$ & $21.47$ & $18.58$ & $SB$ & $1.16$ & $22.43$  & $SB$ & $2.12$ & $23.84$  & $SB$ & $2.71$  \nl
        &      &        &      &        &      &     & & & & & &   \nl
\multicolumn{13}{c}{ } \nl
\enddata
\label{tab:galanal}
\end{deluxetable}
\endgroup
